\documentclass[namedreferences]{kluwer}
\usepackage[dvips]{graphicx}
\usepackage{makeidx}

\makeindex

\newcommand{\FigBox}[2][\columnwidth]{\framebox[#1]{\rule{0pt}{#2}}} 

\begin{document}
\begin{frontmatter}
\end{frontmatter}
\begin{article}
\begin{opening}

\small

\title{Particle modeling of disk-shaped galaxies
 of stars on nowadays concurrent supercomputers}

\author{Evgeny \surname{Griv}\thanks{Corresponding author
 (email: griv@bgumail.bgu.ac.il).}}

\author{Michael \surname{Gedalin}}

\author{Edward \surname{Liverts}}

\author{David \surname{Eichler}}

\institute{Department of Physics, Ben-Gurion University
 of the Negev, P.O. Box 653, Beer-Sheva 84105, Israel}

\author{Yehoshua \surname{Kimhi}}

\institute{High Performance Computing Unit, 
 Inter University Computational Center, Tel Aviv 
 University, Ramat Aviv 69978, Israel}

\author{Chi \surname{Yuan}}

\institute{Academia Sinica Institute of Astronomy and Astrophysics
 (ASIAA), P.O. Box 1-87 Nankang, Taipei 11529, Taiwan} 

\runningtitle{MODELING OF GALAXIES OF STARS}

\runningauthor{E. GRIV ET AL.}

\begin{ao}
 KLUWER~ACADEMIC~PUBLISHERS PrePress Department,\\
 P.O.~Box~17, 3300~AA~~Dordrecht, The~Netherlands\\
 e-mail: TEXHELP@WKAP.NL\\
 Fax: +31 78 6392500
\end{ao}

\begin{abstract}
The time evolution of initially balanced, rapidly rotating models for 
an isolated disk of highly flattened galaxies of stars is calculated.
The method of direct integration of the Newtonian equations of motion 
of stars over a time span of many galactic rotations is
applied.  Use of modern concurrent supercomputers has enabled us 
to make long simulation runs using a sufficiently large number of
particles $N=30,000$.  One of the goals of the present simulation
is to test the validities of a modified local criterion for stability
of Jeans-type gravity perturbations (e.g. those produced by a barlike
structure, a spontaneous perturbation and/or a companion galaxy)
in a self-gravitating, infinitesimally thin and collisionless disk.
In addition to the local criterion we are interested in how model
particles diffuse in velocity.  This is of considerable interest in
the kinetic theory of stellar disks.  Certain astronomical implications 
of the simulations to actual disk-shaped (i.e. rapidly rotating) galaxies
are explored.  The weakly nonlinear, or quasi-linear kinetic theory 
of the Jeans instability in disk galaxies of stars is described as well.
\end{abstract}

\classification{JEL codes}{will be inserted by the editor}


\end{opening}

\section{Introduction}

In galactic astronomy, the most intensely studied of the problems of pattern
formation is the problem of the formation of spiral structure, which has
brought forth numerous theoretical and numerical approaches. 
Even though at the present time there exists,
as yet, no satisfactory theory of the origin and conservation of the spiral
structure so prominent in the Milky Way and many other giant flattened
galaxies, the majority of experts in the field has yielded to the opinion
that the study of the stability of the small-amplitude cooperative
oscillations in disk-shaped galaxies of stars is the first step towards an
understanding of the phenomena.  This is because in the Milky Way and many
other disk galaxies, the bulk of the luminous mass,
probably $\gtrsim 90\%$, is in the stars.  Therefore, it
seems likely that the spiral structures are intimately connected with the
stellar constituent of a galaxy, and that stellar dynamical phenomena play
a basic role.  Because of the long-range nature of the gravitational force
between stars, a galaxy exhibits collective modes of motions --- modes in
which the stars in large regions move coherently or in unison.  A galaxy
is characterized also by a high ``temperature" (a dispersion of random
velocities of stars) and hence a high ``thermal" energy; this thermal 
energy is much larger than the interaction potential between pairs of stars.
Because of this, binary encounters produce only small-scale scattering of
the motions of the stars.  Over long enough times, these gradually build
up to produce large deflections that constitute collisions.  In many cases
these collisional effects are so weak that we consider the galaxy as
collisionless on the Hubble time $T \sim 10^{10}$ yr (Chandrasekhar, 1960; 
Binney and Tremaine, 1987).  Thus, the galaxy is dominated by the 
collective motions and the free streaming of the stars (kinetic effects).

As usual it is very difficult determine a priori whether a particular
linearization of nonlinear equations made in analytical studies 
constitutes a valid approximation.  This can be 
determined, however, by direct computer simulations of the nonlinear 
theory that mimic the behaviour of stars in galaxies.  
Such modeling gives more detailed information than can be
obtained analytically, so that the important physical phenomena can 
be determined.  In this paper, the results of the approximate theoretical
analysis described in the Appendix are compared with our own many-body
($N$-body) simulations.   One of the goals of the simulation is to test
the validities of a modified local criterion for stability of Jeans-type
gravity perturbations in a self-gravitating, infinitesimally
thin and collisionless disk.  The fact that the nonaxisymmetric 
perturbations in the differentially rotating system are more unstable
than the axisymmetric ones is taken into account in this modified
Toomre-like (Toomre, 1964, 1977) criterion.  

In addition to the stability criterion we are interested in how model
particles diffuse in velocity.  As for the present study, such random
velocity diffusion is caused by stellar disk turbulence. 
In particular, we verify the analytical prediction that as the
Jeans-unstable perturbations grow the mean-square velocity increases 
linearly with time.  The observation of this behavior is a sensitive
test of the model as well as the kinetic theory of stellar disks.

In fact, Morozov (1981) and Griv {\it et al.} (1999a) already attempted 
to confirm the modified criterion numerically.  However, because of
the very small number of model stars, $N=200$, Morozov's results are
subject to considerable uncertainties, and additional simulations are
clearly required to settle the issue.   Furthermore, for that number
of model particles, the two-body relaxation time scale is comparable to 
the crossing time, even with Morozov's modest softening parameter, 
raising some question about the applicability of his simulations to 
actual almost collisionless galaxies.  Increasing the number of model 
stars is definitely a more reliable procedure.  In turn, Griv {\it et
al.} used a different numerical approach of the so-called local
$N$-body simulations.  The $N$-body experiments in a local or 
Hill's approximation has been pioneered by Toomre (1990), Toomre
and Kalnajs (1991) and Salo (1995).  
In these simulations dynamics of particles
in small regions of the disk are assumed to be statistically
independent of dynamics of particles in other regions.
The local numerical model thus simulates only a small
part of the system and more distant parts are represented
as copies of the simulated region.  Unlike Morozov (1981) and 
Griv {\it et al.} (1999a), we study some aspects of dynamical 
behavior of stellar systems by global simulations using a 
sufficiently large number of particles.

\section{$N$-body simulations}

Different methods are currently employed to simulate global evolution 
of collisionless point-mass systems of galaxies by $N$-body experiments.
One algorithm for a simulation code that has found wide use was 
developed by Hohl and Hockney (1969), Hohl (1971, 1972), Miller 
(1971), Hockney and Brownrigg
(1974), Quirk (1972), Athanassoula and Sellwood (1986), Combes
{\it et al.} (1990), Pfenniger and Friedli (1993), Merritt and Sellwood
(1994), Donner and Thomasson (1994), and others.  The method samples the
density field with a usually uniform grid.  The Poisson equation is then
solved on this grid using one of a number of the fast solvers that are
available, usually Fast Fourier Transforms.  This is an analog of
plasma particle-in-cell (PIC) codes.  It is believed that simulating 
many billions of stars in actual galaxies by using only several ten 
or hundred thousands representative stars in PIC experiments will be
enough to capture the essential physics, which includes the ordinary
star--star binary relaxation and wave-like collective motion. 
In other fields, such as the simulation of spiral structures, PIC
codes may be used with moderate success.  This is because these
fine-scale $\lesssim 1$ kpc structures can basically be governed
by collisionless processes.  By increasing the number of cells to
reproduce the microstructures, one reduces the average number of
particles per cell, and thus increases the undesirable effect of
particle gravitational encounters.  In addition, if one uses PIC
codes, then one cannot resolve density fluctuations with wavelengths
smaller than the size of a cell $d$.  This puts a lower limit on the
wavelengths that it makes any sense to keep; this lower limit is
given by $\lambda_{\mathrm{min}} \approx d$.  Other disadvantages
of PIC codes are possibilities of late-time loss of energy
conservation, finite-grid and aliasing instabilities, and 
other numerical artifacts.

PIC method works well only for regions where there are
many particles per cell.  The problem is serious because there are
regions in the phase space which are important to the problem but in
which the distribution function is small.  In general, since we will
at most have a hundred particles in one cell we will have fluctuation
of the order of $10\%$ or more which for many problems is unacceptable.

In another type of $N$-body simulation, Ostriker and Peebles (1973),
Morozov (1981), Grivnev (1985), Griv {\it et al.} (1994), Griv and
Zhytnikov (1995), Griv and
Chiueh (1998) investigated a stellar disk by direct integration of
Newtonian equations of motion.  Following them, we simulate here the
evolution of a model for an isolated thin disk of a galaxy by direct
integration of the equations of motion of stars: the force 
on a star $i$ due to all other stars is given by
   \begin{equation}
\frac{d^2 {\bf r}_i}{d t^2} = G
\sum_{j\ne i}^N \frac{m_i m_j ({\bf r}_j -{\bf r}_i)}{[({\bf r}_j
- {\bf r}_i )^2 + r_{\mathrm{cut}}^2 ]^{3/2}} ,
   \end{equation}
where $G$ is the constant of gravitation, $m_i$, $m_j$, ${\bf r}_i$
and ${\bf r}_j$ are the mass, position and velocity of the
$i$-th and of the $j$-th particle, respectively, and $r_{\mathrm{
cut}}$ is the so-called cutoff radius.  This ``softening" of the 
gravitational potential is a device often used in $N$-body
simulations to avoid numerical difficulties at very close 
but rare encounters (Miller, 1971; Romeo, 1998).  
The constant parameter $r_{\mathrm{cut}}$
reduces the interaction at short ranges and puts a lower limit
on the size of the particles, i.e. the stars in the system can no
longer be considered as point-masses --- they are in fact Plummer
spheres with a scale size $r_{\mathrm{cut}}$.  In addition, the
softening parameter $r_{\mathrm{cut}}$ leads to a more
``collisionless" situation.

As is known, the latter $N$-body models suffer from graininess due to 
the finite number of particles.  The graininess manifests itself as
a numerical ``noise" in the calculation.  The use of many model
particles $N \gtrsim 10^4$ decreases drastically the noise level.

Although the direct integration method sounds simple and straighforward,
practical computational limitations require the use of only few ten
thousands representative stars.  This is because we evaluate the sum
on the right-hand side of Eq. (1) for each star or $N$ times for $N$
stars.  The sum itself contains $N$ terms; each term requires a number
of arithmetic operations --- for the sake of estimating, let us say, 
ten.  The total number of arithmetic operations required to evaluate 
the force will be of the order of $10 \times N^2$.  
For the standard Runge--Kutta
calculation involving $30,000$ stars the total number of operations
would be about $4 \times 10^{10}$.  Assuming that we can do an
operation in $10^{-8}$ sec, simply evaluating the forces would
require about $6-7$ min.  A typical calculation requires several
thousand time steps so that $\sim 1,000$ h would be required.
Calculations of this magnitude are totally impractical for using such
models to explore the physics of galaxies.  The problem is solved by
using the nowadays concurrent supercomputers: with the multiple
processors in supercomputers it is possible to reduce the running time
by processing more than one group of particles at a time.  The use of, 
say, $100$ parallel computers would make such calculations practical.
In the present study, we explore the Inter University $112$ 
processors SGI Origin $2000$ supercomputer in Israel.

Note that PIC codes only formally utilized a larger number
of particles than direct integration codes.  This is because in PIC
experiments, the disk plane was covered with a difference grid.  The
stars within grid cells as well as within a whole model did not interact
gravitationally; the computation of the mass density in cells followed a
fast Fourier transform to obtain the potential only at a restricted set
of points --- at the cell centers --- by interpolation rules
and thus to find the very approximate forces at the positions of the
particles.  This avoids the need to compute forces between particle
pairs, and makes the amount of computation necessary to obtain the
forces independent of $N$.  Clearly,
nothing similar exists in nature, and therefore it follows that one
cannot compare the number of particles used in these different kinds of
simulations in any way at all.  Modern parallel computing technology on
concurrent processors had led to a dramatic increase in the computing
power.  The latter allows us to use direct integration methods with
a sufficiently large number of model stars $N=30,000$.

\section{Numerical model}

In general, the numerical procedure is first to seek stationary
solutions to the Boltzmann kinetic equation in the self-consistent
field approximation for the density distribution, the angular velocity 
of regular rotation, the dispersion of random velocities of the stars,
and then to calculate the stability of those solutions to 
small-amplitude gravity perturbations.

At the start of the $N$-body integration, our simulation initializes
the particles on a set of $100$ circular rings with a circular
velocity $V_{\mathrm{rot}}$ of galactic rotation in the $r, \,
\varphi$--plane; the system is isolated in vacuum.  Consider an 
uniformly rotating model disk of stars with a surface mass density
variation given by
   \begin{equation}
\sigma_0 (r) = \sigma (0) \sqrt{1 - r^2 / R^2} ,
   \end{equation}
where $\sigma (0)$ is the central surface density and $R$ is the radius
of the initial disk.  As a solution of a time-independent collisionless
Boltzmann equation, to ensure initial equilibrium the uniform angular
velocity to balance the zero-velocity-dispersion disk,
   \begin{equation}
\Omega_0 = \pi \sqrt{G \sigma (0) / 2 R} ,
   \end{equation}
was adopted (Hohl, 1971, 1972).
Then the position of each particle was slightly perturbed by applying
a pseudorandom number generator.  For the uniformly rotating disk,
the Maxwellian-distributed random velocities with radial $c_r$ and
azimuthal $c_\varphi$ dispersions in the plane $z=0$ according to
the familiar Toomre's criterion (Toomre, 1964, 1977) ,
   \begin{equation}
c_{\mathrm{T}} = \frac{3.36 \sigma_0}{\kappa}
= 0.341 \Omega_0 \sqrt{R^2 - r^2} ,
   \end{equation}
may be added ($c_r = c_\varphi$) to the initial circular velocities
$V_{\mathrm{rot}} = r \Omega_0$, where $\kappa = 2 \Omega [1+
(r/\Omega)(d\Omega /dr) ]^{1/2}$
is the ordinary epicyclic frequency (Hohl, 1971, 1972).  It is crucial
to realize that in this case, according to Lau and Bertin (1978), Lin 
and Lau (1979), Bertin (1980), Lin and Bertin (1984), Morozov (1980),
Polyachenko (1989),
Griv and Peter (1996), Polyachenko and Polyachenko (1997), 
initially the disk is Jeans-stable against the small-scale axisymmetric
(radial) perturbations but unstable against the relatively large-scale
nonaxisymmetric (spiral) perturbations.  The initial vertical velocity
dispersion was chosen equal to $c_z = 0.2 c_r$.  Finally, the angular
velocity $\Omega_0$ was replaced by (Hohl, 1971, 1972)
   \begin{displaymath}
\Omega = \left\{ \Omega_0^2 + \frac{1}{r \sigma (r)}
\frac{\partial}{\partial r} \left[ \sigma (r) c_r^2 (r)
\right] \right\}^{1/2} .
   \end{displaymath}

The sense of disk rotation was taken to be counterclockwise, and units
are such that $G=1$ and the mass of a star $m_{\mathrm{s}} = 1$. 
The cutoff radius was chosen $r_{\mathrm{cut}} = 0.01 R$.
Within the framework of our model, dynamically young stars
form the very thin disk $h \ll R$.  Therefore, at the start
of simulations the disk thickness was chosen $h=0.05 R$.
A time $t=1$ is taken to correspond to a single revolution of the
initial disk.  In all experiments the simulation was performed
up to a time $t$ between $8$ and $10$.
It should be noted here that after about three rotations the picture
is always stabilized and practically no significant changes in gross
properties of the model over this time are observed.  Any difference between
the results of simulations with and without applying the so-called quiet
starts procedure to select the very regular initial coordinates of particles
was not found.  (Basically, by applying the method of quiet starts,
one uses no random numbers in the initial conditions to suppress the
noise level in a system.  Such techniques have proven useful in
obtaining realistic noise levels without the use of a large number
of particles; Byers and Grewal, 1970.)
Moreover, tests indicated that the results were insensitive to changes
in other parameters, e.g. the number of stars in the range $N=10,000
-40,000$, the cutoff parameter in the range $r_{\mathrm{cut}} = (0.005-
0.05) R$ and the initial disk thickness in the range $h = (0-0.2) R$.
Spiral amplitudes detected in our simulations are greatly in excess of 
an expectation from the level of particle noise and do not scale with 
$N$.  This behavior is clearly inconsistent with the hypothesis that
the nonaxisymmetric structure in the models can be attributed to
collisional effects, inappropriate values of $r_{\mathrm{cut}}$
or swing-amplified particle noise as advocated by Toomre (1990).

Finally,
slight corrections have been applied to the resultant velocities and
coordinates of the model stars so as to ensure the equilibrium between
the centrifugal and gravitational forces, to preserve the position of
the disk center of gravity at the origin and to include the weak effect
of the finite thickness of the disk to the gravitational potential.  Thus,
the initial model is very near the dynamical equilibrium for all radii.

The system of equations of motion (1)
for $N$ particles was integrated by the standard Runge--Kutta method
of the fourth order.  All the particles move with the same constant
Runge--Kutta time step $\Delta t = 0.001 t_{\mathrm{orb}}$, where
$t_{\mathrm{orb}}$ is the orbital period.

\section{Results}

\subsection{Collisions}

In order to be compared with an actual galaxy of stars, the
numerical model of the disk should properly simulate a collisionless
system.  Otherwise, as it has been shown, e.g. by Morozov {\it et al.}
(1985) and Griv {\it et al.} (2000a),
a collisional disk is unstable with respect to a secular
dissipative-type instability.  The secular instability of spontaneous
gravity perturbations might arise merely from the dissipation of
the energy of regular rotation into ever larger amounts of heat when
the system has moderate dissipation; it results from perturbations
which have negative energy in the dissipative medium (Morozov {\it
et al.}, 1985; Fridman and Polyachenko, 1984, Vol. 2, p. 243).  
The secular instability produces structures completely unrelated 
to the effects we would like to model.
In the spirit of the simple ``molecular-kinetic" theory
by Chandrasekhar (1960), the collisional relaxation time for
a three-dimensional system from the point of view of deflections
suffered by a test star in encounters with a given distribution
of field objects is
   \begin{equation}
\tau \approx \frac{c^3}{2 \pi G^2 
m_{\mathrm{f}}^2 n_{\mathrm{f}} \ln \Lambda_N } ,
   \end{equation}
where $c$ is the averaged velocity dispersion, 
$m_{\mathrm{f}}$ is the ``field" particle mass and $n_{\mathrm{f}}$
its three-dimensional number density.
Also $\ln \Lambda_N$ is the so-called Newton (or Coulomb in plasmas)
logarithm, by means of which the long-range nature of the gravitational
force is taken into account; to the order of magnitude 
$\ln \Lambda_N \sim \ln N_{\mathrm{f}}$, 
where $N_{\mathrm{f}}$ is the total number
of field particles.  In actual galaxies
$\ln \Lambda_N \approx 20$.  According to Eq.
(5), the chosen quantities $c$, $N$, etc. guarantee that the model
may be considered as a collisionless one to a good approximation at
least during $\sim 50$ rotations.  So we expect, in our numerical
calculations because $N$ is large, relaxation from two-body
gravitational interactions is unimportant.  We
suggest that the structures observed in the $N$-body simulations
originate from the collective modes of practically collisionless
galactic models --- gravitational Jeans-type modes and bending
firehose-type modes (see below Section 4.2). 

Consider a system of mutually gravitating particles.  The local
distribution function $f ({\bf r},{\bf v},t)$ must satisfy
the Boltzmann kinetic equation
   \begin{equation} 
\frac{\partial f}{\partial t} + {\bf v \cdot} \frac{\partial f}{ 
\partial {\bf r}} - \nabla \Phi {\bf \cdot}
\frac{\partial f}{\partial {\bf v}} = \left( \frac{\partial f}{ 
\partial t} \right)_{\mathrm{coll}} ,
   \end{equation}
where $\Phi ({\bf r},t)$ is the total gravitational potential
determined self-consistently from the Poisson equation, 
   \begin{equation}
\nabla^2 \Phi = 4 \pi G \int f d {\bf v} ,
   \end{equation}
$(\partial f / \partial t)_{\mathrm{coll}} \propto \nu_{\mathrm{c}}
(f_0 - f)$ is the so-called collisional integral which defines the
change of $f$ arising from ordinary interparticle collisions,
$\nu_{\mathrm{c}}$ is the collision frequency and $f_0$ is the
quasi-steady state distribution function (Binney and Tremaine, 
1987, p.~506; Griv {\it et al.}, 2000a).

In plasma physics,
Lifshitz and Pitaevskii (1981, p. 115) have discussed
phenomena in which interparticle collisions are unimportant, and such a
plasma is said to be collisionless (and in the lowest-order approximation
of the theory one can neglect the collision integral in the kinetic
equation).  It was shown that a necessary condition is that
$\nu_{\mathrm{c}} \ll |\omega|$, where $\omega$ is the typical
frequency of collective oscillations: 
then the collision operator in the kinetic equation is small
in comparison with $\partial f / \partial t$.  In the Appendix, 
we have shown that the frequency of Jeans oscillations in
a stellar disk $\omega \sim \Omega$.  Therefore, in the gravitation case
in the lowest-order approximation of the theory we can neglect the effects
of collisions between particles on a time scale of many rotations if $\nu_c
\ll \Omega$.  Lifshitz and Pitaevskii 
have pointed out that collisions may be neglected also if the particle
mean free path is large compared with the wavelength of collective
oscillations.  Then the collision integral in Eq. (6) is small in
comparison with the term ${\bf v} \cdot (\partial f / \partial {\bf r})$.

In this subsection, we test numerically if the models used in our $N$-body
simulations are being correctly modeled as collisionless Boltzmann
(Vlasov) systems.  The direct method of checking if the system is being
modeled as a collisionless system is to repeat a calculation using
a mass spectrum (Rybicki, 1971).  It is obvious that as a result of
gravitational collisions there is a tendency towards energy equipartition
between the various masses.  Hohl (1973) has determined the experimental
relaxation time and compared it with a theoretical prediction for the
collisional relaxation time of a two-dimensional disk.  Here we do such
a comparison by using the three-dimensional system.

Let us consider the three-dimensional computer model consisting of
$2\%$ stars of mass $m_3 = 10 m_{\mathrm{s}}$, 
$18\%$ stars of mass $m_2 = 2 m_{\mathrm{s}}$
and $80\%$ stars of mass $m_1 = 0.55 m_{\mathrm{s}}$.  
The total number of
stars, which are distributed in the system in accord with the law
(2) is $N=20,000$.  Initially, the different mass groups of stars
are distributed with the same velocity dispersion (with different
temperatures).

In the Appendix of the present paper we demonstrate that the 
disk becomes almost stable gravitationally for $c_r \gtrsim
2c_{\mathrm{T}}$.  In such a Jeans-stable system collective 
effects associated with the classical gravitational instability will
not affect the random velocity dispersion of particles: 
then the change of velocity 
dispersion can be explained only by usual two-body encounters.  For
this reason the initial condition was chosen to be a quasi-stable
uniformly rotating disk with $c_r = 2c_{\mathrm{T}}$.  
Following Chandrasekhar (1960) and
Hohl (1973), let us define the relaxation time $\tau_{\mathrm{E}}$ 
as the time required for the mean change of the kinetic energy per 
unit mass of the test star to equal the initial kinetic energy.

In Figure 1 we show the change of the ratio of the mean particle
(kinetic) energy, $< m_1 V_1^2 >$, $< m_2 V_2^2 >$ and $< m_3 V_3^2 >$
(in units of the total kinetic energy of the system), for the different
mass groups, where $V_i$ is the total velocity of a given mass group.
As is expected, the two groups of heavy stars lose energy while the
group of lightest stars gains an approximately corresponding amount of
kinetic energy.  Also as is expected, one can see the decrease in the
change of the kinetic energy with time.  This is because the
collisional frequency $\nu_c \approx 1 /\tau_{\mathrm{E}}$ is 
inversely proportional to the velocity dispersion (Eq. [5]),
and thus the encounters only weakly affect the stars with high
random velocities.

\begin{figure}
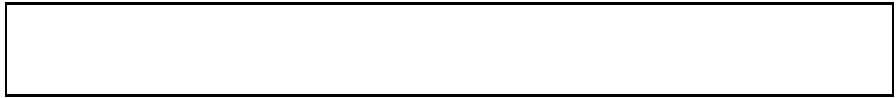

\FigBox{1cm}
\caption{Rate of change of the mean kinetic energy for stars of the
three mass groups of the Jeans-stable model with $N = 20, 000$ and
$c_r = 2c_{\mathrm{T}}$; 1 --- kinetic energy of stars with the mass
of a star $10 m_{\mathrm{s}}$, 2 --- with the mass of a star $2 m_
{\mathrm{s}}$ and 3 --- with the mass of a star $0.55m_{\mathrm{s}}$.
The two groups of heavy stars lose kinetic energy while the
group of lightest stars gains an approximately corresponding
amount of kinetic energy.  In general agreement with the 
analytical estimation, the mean slope of the curves will result
in energy equipartition after about $100-200$ rotation periods.
This result shows that the model used in our $N$-body
simulations is ``collisionless" to a good approximation
and suggests that interparticle collisions do not play
a significant role for instabilities studied in the paper.}
\label{grivetal1.eps}
\end{figure}

As one can see, the mean slope of the curves shown in Figure 1 will
result in energy equipartition after about $100-200$ rotation periods.
It is crucial to realize that these relaxation times even for this
relatively small number of model stars are much longer that the time 
of a single disk revolution.  We conclude that the computer models
used in the present study may be considered as collisionless
ones to a good approximation at least during the first $10$
rotations which are of especial interest in spiral-galaxy simulation.
Therefore, we argue that the collective effects studied in the
paper were apparent before the collisional time scale associated 
with the binary interactions of stars was reached.

\subsection{Warm Toomre-stable disk}

To study the development of spiral structure and to confirm
the theoretical predictions made in the Appendix, warm disks with 
initial radial velocity dispersion of identical stars
equal to Toomre's (1964)
stabilizing velocity dispersion $c_{\mathrm{T}}$ were modeled.
Figure 2 shows an example of the development of spiral structures.
The structure grows rapidly on a dynamical time scale $\sim \Omega
^{-1}$.  At a time $t \approx 0.6$ a multi-armed filamentary spiral
structure forms with several short, tightly wound trailing 
arms.\footnote{Interestingly, according to fairly recent observations,
such ragged galaxies with almost chaotic-looking, ``flocculent" spiral
structures are more common than the grand design ones (Binney and
Tremaine, 1987, p.~391).}  Soon after, at $t \approx 2.8$ a bar
structure appears with two almost symmetrical open spiral arms.
However, sometimes, e.g. at $t=1.6$ or at $t=2.6$, 
one can see a ``one-armed"
galaxy consisting of a prominent arm and a short, diffuse second
one.  After $\sim 3$ rotations, the spiral arms almost disappear,
and a bar forms in the central parts of the system. 

\begin{subfigure}[alph]
\begin{figure}
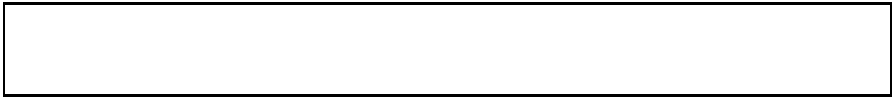

\FigBox{1cm}
\caption{The time evolution of a warm Toomre-stable three-dimensional 
disk of stars.  The system is violently unstable to low-$m$ spiral
Jeans-type modes.  After $\sim 3$ rotations the spiral structure
almost dissapears, and a prominent bar forms.} 
\label{grivetal2a.eps}
\end{figure}

\begin{figure}
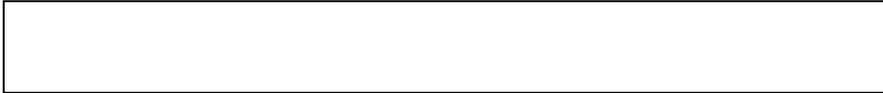

\FigBox{1cm}
\caption{{\it Continued}}
\label{grivetal2b.eps}
\end{figure}
\end{subfigure}

The $m=1$ instability
shifts the point with highest density from the center of mass.  It is
interesting to note that in many disk-shaped galaxies, e.g. in the
spiral galaxies M 101 and NGC 1300, 
there appears to be a deviation from rotational
symmetry.  In principle, such a deviation may be due to this one arm
instability.  Note that the latter has already been suggested by B.
Lindblad (see Toomre, 1996).  In fact, asymmetries in the distribution
of light in spiral galaxies have been known for a long time.
Baldwin {\it et al.} (1980) presented a number of spiral galaxies
in which the distribution of neutral gas is lopsided, i.e. the gas
extends further out on one side of the galaxy than on the other.
In many systems asymmetries are also found to be well developed in
the stellar disks  as indeed confirmed recently by near-infrared
observations (Zaritsky and Rix, 1997).  Rudnick and Rix (1998)
quantified the mean asymmetry of 54 early-type disk galaxies using
the amplitude of the $m=1$ azimuthal Fourier component of the
$R$-band surface brightness.  It was found that $20\%$ of all disk
galaxies have azimuthal asymmetries in their stellar disk 
disribution, confirming lopsidedness as a dynamical phenomenon.
Based on the $N$-body simulations, we can offer a tentative 
hypothesis which accounts lopsided asymmetries in spiral galaxies 
by the $m=1$ Jeans-unstable spiral mode.

Following Athanassoula and Sellwood (1986), in Figure 3 the time
evolution of the Fourier spectrum of the spiral pattern shown in
Figure 2 is plotted.  Only the azimuthal $m=1-6$ components
are shown.  The complex Fourier coefficients, $A$, are determined
from the following summation
   \begin{displaymath}
A (p,m) = {1 \over N} \sum_{j=1}^N \exp \{ i [ m \varphi_j
+ p \ln (r_j) ] \} ,
   \end{displaymath}
where $N$ is the number of particles and $(\varphi_j,r_j)$ are
the polar coordinates of the $j$-th particle.  The pitch angle of
an $m$-armed logarithmic spiral $\psi$ is given by $\psi = \arctan
(p/m)$, and positive $p$ corresponds to trailing spirals and 
negative $p$ to leading ones.

\begin{figure}
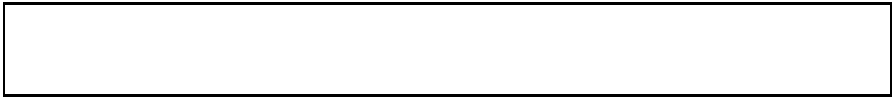

\FigBox{1cm}
\caption{The time evolution of the Fourier spectrum of the particle
distribution shown in Figure 2 for different azimuthal mode numbes  
$m$.  In this figure, the ordinate is the
Fourier coefficient $A(p,m)$ and the abscissa is the pitch angle
$p$.  Spiral amplitudes are greatly in 
excess of an expectation from the level of particle noise $\sqrt{\pi
/4N}$; true instabilities develop in the model. The result suggests that 
very flat stellar disks of galaxies can be unstable to the most unstable
one-armed ($m=1$), two-armed ($m=2$) and three-armed ($m=3$) spiral
Jeans-type gravity perturbations.  The one-armed mode may explain the
asymmetric patterns observed in many isolated galaxies.}
\label{grivetal3.eps}
\end{figure}

This logarithmic spiral Fourier analysis shows that the peaks 
of the signals move to positive $p$ with increasing time, reach
maxima at $t = 1.0-1.4$, and eventually decay. It follows, then, that
after a time $t = 2.0-3.0$ the model becomes practically stable 
with respect to gravitational Jeans-type modes.  We can observe
that the $m=1-3$ modes are the most unstable ones.  
An interesting feature is that all these modes have roughly the 
same pitch angle.  The amplitudes of the signals grow aperiodically,
supporting the assumption that we have deal with the aperiodic
Jeans-type instability of gravity disturbances.

In actual galaxies, discrete spiral modes were 
already found by Rix and Zaritsky (1995), Block and 
Puerari (1999) and Block {\it et al.} (1999).  It was stated that
in the near-infrared, the morphology of older star-dominated disk 
indicates a simple classification scheme: the dominant Fourier 
$m$-mode in the dust penetrated regime, and the associated pitch
angle.  A ubiquity of low $m=1$ and $m=2$ modes was confirmed. 
Fridman {\it et al.} (1998) detected $m=1-8$ spiral modes in
relatively young stellar population of the nearly face-on galaxy 
NGC 3631 from observations in the H$_\alpha$ line. 

From the edge-on view pictured in Figure 4, one can see that
from an initial very thin model, a fully three-dimensional disk
develops immediately at $t \approx 0.4$ with a mean height $<z>$ 
above the plane, corresponding to the 
force balance between the gravitational attraction in the plane and 
the ``pressure" due to the velocity dispersion (i.e. ``temperature")
in the vertical, $z$-direction.  A straightforward estimate shows
that
   \begin{equation}
<z> \approx \frac{c_z}{\sqrt{4 \pi G \mu}} ,
   \end{equation}
where $\mu$ is the three-dimensional mass density in the
mid-plane.  After a time
$t \approx 0.4$ there is no change in the edge-on structure
until at $t \approx 1.4$ (when the Jeans instability is almost 
switched off by increasing the planar velocity dispersion).
At somewhat later times, $t \approx 3.2$, a ``box-shaped" or
``peanut-shaped" bar structure is developed.  The development
of the box/peanut bar structure is explained in the terms of the
bending firehose-type instability (Raha {\it et al.}, 1991;
Merritt and Sellwood, 1994; Griv and Chiueh, 1998).

\begin{subfigure}[alph]
\begin{figure}
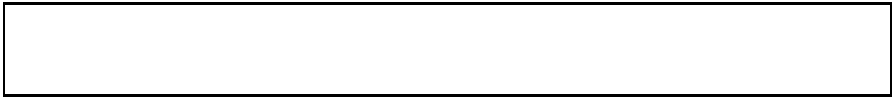

\FigBox{1cm}
\caption{Higher-resolution plots of the central parts (edge-on
view) for the simulation run seen in Figure 2.  At a time $t \approx
1.4$, one can see the rapid development of the snake-shaped, bending
firehose-type instability as a precursor of galactic bulge
formation in the central, rigidly rotating portions of the system. 
This instability is a transient one, and develops
on a time scale of single rotation of the system only.
The linear dimension of the bending along the $z$-axis
is comparable to the disk thickness and much less than
the radius of the outermost parts of the system.  
The usual name for this instability of collective
systems --- firehose instability --- recalls the fact that 
the unstable motion of a hose when the flow velocity of water 
inside becomes very high has essentially the same nature.}
\label{grivetal4a.eps}
\end{figure}

\begin{figure}
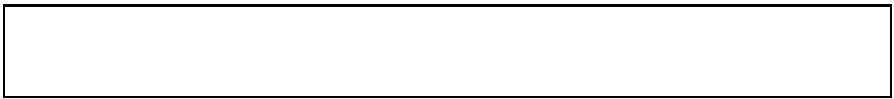

\FigBox{1cm}
\caption{{\it Continued}}
\label{grivetal4b.eps}
\end{figure}
\end{subfigure}

One sees that at times
$t = 1.4-2.4$, the bending instability fiercely develops in the
central, almost rigidly rotating parts of the system and is switched off
at $t \approx 2.8$.  At later times, no dramatic evolution is observed
in our simulations.  To emphasize, this type of bending instabilities
develops in the central, rigidly rotating parts of the system.  The
three-dimensional simulations show that the bending instabilities
mentioned above destroy the strong bar formation (Griv and Chiueh,
1998).  Note that the bending modes have been found experimentally
by Merritt and Hernquist (1991) and
Merritt and Sellwood (1994), but in the nonrotating models.  Combes
{\it et al.} (1990) and Raha {\it et al.} (1991) found the bending
instability in the central regions of a fast-rotating $N$-body disk.
It was stated that this instability may play a part in the formation
of triaxial bulges in flat galaxies and explanation was given why
many bulge stars are less than 10 Gyr old (Raha {\it et al.}, 1991).
Bureau and Freeman (1997) have argued that new observations 
constitute a strong case in favor of the bending (or bar-buckling)
mechanism for the formation of boxy/peanut-shaped bulges in spiral
galaxies.  Observations suggest very strongly that many of ordinary
spiral galaxies, including the Milky Way, have similar small 
triaxial bars in their central regions (Dwek {\it et al.}, 1995).

Apparently, this instability of a sufficiently
thin stellar disk has been predicted by Toomre (1966) by using the
simplified theory based on moment equations.  Toomre considered the
collisionless analog of the Kelvin--Helmholtz instability in an infinite,
two-dimensional, nonrotating sheet of stars.  This is the
usual way to discuss the conditions of the firehose instability 
in plasma physics (Krall and Trivelpiece, 1986).  It was
demonstrated by Toomre that the bending instability is driven
by the stellar ``pressure" anisotropy: the source of free energy in
the instability is the intrinsic anisotropy of a velocity dispersion
(``temperature").  The bending firehose-type instability was also
discovered independently by Kulsrud {\it et al.} (1971) and Mark 
(1971) with a more accurate kinetic theory.
Kulsrud {\it et al.} have described the
mechanism of the instability in terms of the balance of centrifugal force
acting on the stars of a bending layer and the gravitational attraction
toward the plane of the practically nonrotating, pressure-supported disk.
Fridman and Polyachenko (1984) have discussed the role
of the instability in explaining the existence of maximum oblateness
in almost nonrotating elliptical galaxies and the formation of the
bulges of disk-shaped galaxies of stars.

As is seen in Figures 2 and 4, the basic form of warm disk evolution
is the expansion of the outermost parts together with the contraction of
a large part of the mass towards the center; in the final quasi-steady
state the surface density decreases exponentially outwards (Fig. 5).
In a future publication we intend to explain the exponential surface
density distribution of the final quasi-steady state model
by applying the quasi-linear theory of a stellar disk's stability.
Note only that the study of surface photometry have been shown that
most spiral and S0 galaxies have an exponential disk with radial
surface-brightness distribution $\sim \exp ( - \mbox{const} \cdot
r)$ (Freeman, 1970).

\begin{figure}
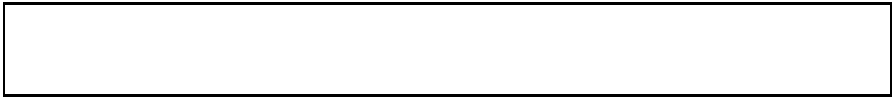

\FigBox{1cm}
\caption{Variation of the surface density for a disk shown in Figure
2.  Note the rapid change with time of the surface density at $t <
1$.  The basic form of the disk evolution is the expansion of the
outermost portions together with the contraction of a large part of
the mass towards the center. In a quasi-steady state after about $2 
-3$ rotations, the surface density decreases exponentially outwards.
According to observations, stellar disks of most spiral and S0
galaxies are similar approximated well by a function that is
exponential in $r$.}
\label{grivetal5.eps}
\end{figure}

At first, the surface density and rotation velocity change
rapidly (Figs. 5 and 6), but after a time $t \sim 1.0$ there is
little further change during the following two rotations,
except in the central portions of the disk.  The model generates
a nonflat rotation curve (Fig. 6); in addition growing density waves
(spiral arms) transfer the angular momentum of the circular motion
outwards (see the difference in the rotation curves at
$t=1.0$ and $t=2.0$ in Figure 6).  As mentioned above,
the model practically shows no the time evolution of the surface
density and the velocity curve after $t \approx 2$.
The latter seem to indicate that at that time the system has
nearly reached a quasi-steady state (cf. Hohl, 1971).

\begin{figure}
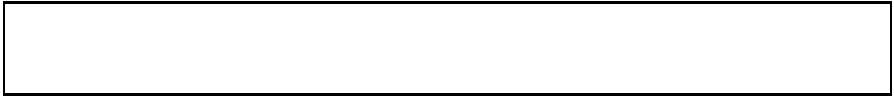

\FigBox{1cm}
\caption{The azimuthal average rotation curve for a stellar disk
shown in Figure 2.  Note the non-flat rotation curve at the 
periphery of the system; growing density waves (spiral arms) 
transfer the angular momentum of the circular motion outwards.}
\label{grivetal6.eps}
\end{figure}

The time evolution of the velocity dispersions $c_r$, $c_\varphi$
and $c_z$ is shown in Figure 7.

\begin{figure}
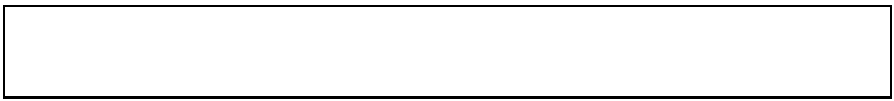

\FigBox{1cm}
\caption{Variation of velocity dispersions for a disk shown in
Figure 2; 1 --- radial velocity dispersion, 2 --- azimuthal velocity
dispersion and 3 --- vertical velocity dispersion.
Initially, during the first rotation the velocity
dispersions increase with time rapidly; later the dispersions
grow only slightly.  However, in the central portions of the disk the 
vertical velocity dispersion $c_z$ grows rapidly even after the time
$t \approx 1.5$, when the bending firehose-type instability begins to
to operate (see Figure 4).  In this central region the bending 
instability strongly changes the ratio of vertical to radial (and 
tangential) dispersion by transfering kinetic energy of plane random 
motions of particles to vertical motions.}
\label{grivetal7.eps}
\end{figure}

Initially, during the first rotation the velocity dispersions
along each axis are changed rapidly.  During the following two rotations
the velocity dispersions are changed with time slightly, confirming
that between $t=2$ and $t=3$ the system has nearly reached a quasi-steady
state.  However, in the central portions of the disk the velocity
dispersion $c_z$ increases rapidly even after the time $t \approx 1.5$, 
when the bending instability begins to operate. 

As was stated, one of the goals of the present research is to compare
Toomre's (1964) critical dispersion (4) and the modified dispersion given
by Eq. (25) with the experimental radial velocity dispersion of the 
quasi-steady state model of a stellar disk.  In what follows, we are
doing such a comparison. 

\begin{figure}
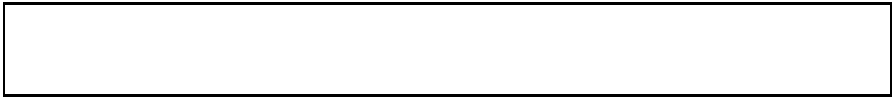

\FigBox{1cm}
\caption{Comparison of Toomre's radial-velocity dispersion and the
modified critical dispersion with the experimental radial velocity
dispersion of a disk shown in Figure 2 at $t = 2.5$;
1 --- the experimental radial velocity dispersion,
2 --- the modified velocity dispersion,
3 --- Toomre's velocity dispersion.} 
\label{grivetal8.eps}
\end{figure}

As is seen in Figure 8, the modified critical dispersion, in contrast
to Toomre's (1964) critical dispersion, is in fair agreement with
the experimental data, especially for the outer, differentially
rotating parts of the disk.  The agreement is quite reasonable
considering the rough nature of the theory.  In the central parts of
the model the agreement with the theory is not good.  This is to be
expected since the condition of nearly circular 
orbits adopted in the Appendix does not fulfill in the central
almost nonrotating regions of model disks (and in spiral galaxies, 
including both barred and normal galaxies, and in disks of
elliptical galaxies).  Therefore in the central parts of the disk 
($r \lesssim 0.4-0.5$) the condition for the local criteria (both
the Toomre and the modified criterion) to be applicable is violated.

One of the important problems of stellar disks is the determination
of the random velocity diffusion.  Such velocity diffusion can be
caused by stellar disk turbulence.  Velocity diffusion produced by
turbulence tells us something about the turbulence process.  

To compute velocity diffusion we calculate the mean-square spread in
the planar random (residual) velocity $v_\perp^2=c_r^2 +c_\varphi^2$ 
as a function of time of model particles for different radii $r$. 
In general $v_\perp^2$ has a time dependence like that shown in 
Figure 9.

\begin{figure}
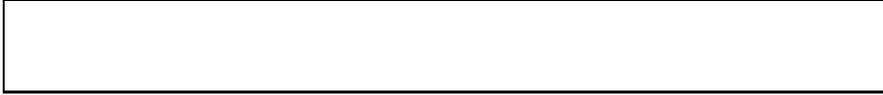

\FigBox{1cm}
\caption{The time development of the mean-square spread in planar
random velocity $v_\perp^2$ of model particles for different radii
$r$.}
\label{grivetal9.eps}
\end{figure}

As is seen, in the differentially rotating ($r > 0.4-0.5$), 
Jeans-unstable (at times $t \lesssim 1.8-2.0$) parts of the disk
approximately $v_\perp^2 \propto t$.  This 
behavior has been predicted analytically (see the Appendix). 
Interestingly, observations indicate a similar increase of 
$v_\perp^2$ with time in the solar neighborhood (Binney and 
Tremaine, 1987, p. 470; Gilmore {\it et al.}, 1990).

\subsection{Hot Jeans-stable disk}

The time evolution of the hot disk was investigated, in which
the initial dispersion of random radial velocities of model
stars was chosen to be greater than Toomre's critical dispersion,
namely, $c_r = 2 c_{\mathrm{T}}$ (see Eq. [24]).  For such 
calculations we obtain plots like those shown in Figure 10.  
In agreement with the theory, the system becomes gravitationally
(Jeans-) stable.

\begin{figure}
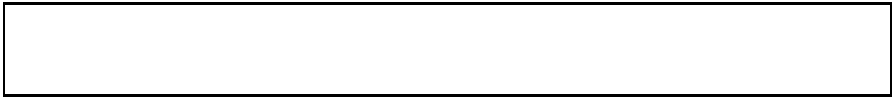

\FigBox{1cm}
\caption{The time evolution of a hot three-dimensional
disk of stars.  The system is stable to all Jeans modes.
The result agrees with the theoretical explanation  
described in the paper.}
\label{grivetal10.eps}
\end{figure}

\section{Appendices}

The linear Lin--Shu density wave theory is considered now as well
established (Lin and Shu, 1966; Lin {\it et al.}, 1969; Shu,
1970; Bertin, 1980; Binney and Tremaine, 1987, p. 347;
Griv and Peter, 1996; Griv 
{\it et al.}, 2000b).  However, as soon as the amplitude of the 
gravity perturbations becomes appreciable a new generation 
of problems, namely, nonlinear problems, become important.
Below we present a quantitative weakly nonlinear theory of
the classical Jeans instability of small-amplitude gravity 
disturbances (e.g. those produced by a barlike structure, 
a spontaneous perturbation and/or a companion galaxy) in
self-gravitating, rapidly rotating stellar disks of flat galaxies.

In the rotating frame of a disk galaxy, the collisionless motion
of an ensemble of identical stars in the plane of the system can
be described by the Boltzmann equation for the distribution
function $f ({\bf r},{\bf v},t)$ without the integral of
collisions (Lin {\it et al.}, 1969):
   \begin{eqnarray}
\frac{\partial f}{\partial t} + v_r \frac{\partial f}{\partial
r} + \left( \Omega + \frac{v_\varphi}{r} \right)
\frac{\partial f}{\partial \varphi}
+ \left(2\Omega v_\varphi + \frac{v_\varphi^2}{r}
+ r \Omega^2 - \frac{\partial \Phi}{\partial r} \right)
\frac{\partial f}{\partial v_r} \nonumber \\
- \left( \frac{\kappa^2}{2\Omega} v_r + \frac{v_r v_\varphi}
{r} + \frac{1}{r} \frac{\partial \Phi}{\partial
\varphi}\right) \frac{\partial f}{\partial v_\varphi} = 0 ,
\label{eq:kinetic}
   \end{eqnarray}
where $r,\varphi, z$ are the galactocentric cylindrical coordinates,
the total azimuthal velocity of the stars was represented as a sum of
the random $v_\varphi$ and the basic rotation velocity
$V_{\mathrm{rot}} = r\Omega$, and $v_r$ is the random velocity in the 
radial direction.
The quantity $\Omega(r)$ denotes the angular velocity of galactic
rotation at the distance $r$ from the center, and 
the epicyclic frequency $\kappa$ varies
from $2\Omega$ for a rigid rotation to $\Omega$ for a Keplerian
one.  Random velocities are small compared with $r\Omega$.
Collisions are neglected here because the collision frequency is
much smaller than the cyclic frequency.  In
the kinetic equation (9), $\Phi ({\bf r},t)$ is the
total gravitational potential determined self-consistently from the
Poisson equation (7).

The equilibrium state is assumed, for simplicity, to be 
an axisymmetric and spatially homogeneous stellar disk. 
Secondly, in our simplified model, the perturbation is propagating in
the plane of the disk.  This approximation of an infinitesimally thin
disk is a valid approximation if one considers perturbations with a
radial wavelength that is greater than the typical disk thickness.  We
assume here that the stars move in the disk plane so that $v_z=0$.
This allows us to use the two-dimensional distribution function $f=f
(v_r,v_\varphi,t)\delta(z)$ such that $\int f dv_r dv_\varphi
dz=\sigma$, where $\sigma$ is the surface density.  We expect that the
waves propagating in the disk plane have the greatest influence on the
development of structures in the disk.  The latter suggestion is
strongly supported by numerical simulations (Hohl, 1978).

The disk in the equlibrium is described by the following equation:
   \begin{equation}
\left( 2\Omega v_\varphi + \frac{v_\varphi^2}{r} \right)
\frac{\partial f_{\mathrm{e}}}{\partial v_r}
- \left( \frac{\kappa^2}{2\Omega} v_r + \frac{v_r v_\varphi}{r}
\right) \frac{\partial f_{\mathrm{e}}}
{\partial v_\varphi} = 0 , \label{eq:equil}
  \end{equation}
where $\partial f_{\mathrm{e}}/\partial t=0$ and the angular
velocity of rotation $\Omega (r)$ is such that the necessary
centrifugal acceleration is exactly provided by the central
gravitational force $r\Omega^2=\partial \Phi_{\mathrm{e}}/\partial r$,
where $\Phi_{\mathrm{e}}$ is the axisymmetric equilibrium potential.
Equation (10) does not determine the equilibrium distribution 
$f_{\mathrm{e}}$ uniquely.
For the present analysis we choose $f_{\mathrm{e}}$ in the
form of the anisotropic Maxwellian (Schwartzschild) distribution
  \begin{equation}
f_{\mathrm{e}} = \frac{\sigma_{\mathrm{e}}}{2\pi c_r c_\varphi}
\exp \left( - \frac{v_r^2}{2c_r^2}
- \frac{v_\varphi^2}{2c_\varphi^2} \right)
= \frac{2\Omega}{\kappa} \frac{\sigma_{\mathrm{e}}}{2\pi c_r^2}
\exp \left( - \frac{v_\perp^2}{2c_r^2} \right) . \label{eq:maxwell}
   \end{equation}
The Schwarzschild distribution function is a function of the two
epicyclic constants of motion ${\mathcal{E}}=v_\perp^2/2$ and
$r_0^2\Omega
(r_0)$, where $r_0 = r + (2\Omega /\kappa^2) v_\varphi$.  These
constants of motion are related to the unperturbed star orbits:
  \begin{eqnarray}
&& r = - \frac{v_\perp}{\kappa} \left[ \sin (\phi_0 - \kappa t)
- \sin \phi_0 \right] ; \quad
v_r = v_\perp \cos \left( \phi_0 - \kappa t \right) ; \nonumber \\
&& \varphi = \frac{2\Omega}{\kappa} \frac{v_\perp}{r_0 \kappa}
\left[ \cos (\phi_0 - \kappa t) - \cos \phi_0 \right] ; \nonumber \\
&& v_\varphi \approx
r_0 \frac{d\varphi}{dt} + r_0 \frac{v_\perp}{\kappa}\frac{d\Omega}
{dr} \sin \left( \phi_0 - \kappa t \right)
\approx \frac{\kappa}{2\Omega} v_\perp
\sin \left( \phi_0 - \kappa t \right) , \label{eq:orbits}
   \end{eqnarray}
where $v_\perp$, $\phi_0$ are constants of integration, $v_\perp
/ \kappa r_0 \sim \rho /r_0 \ll 1$, $\rho$ is the mean epicycle
radius, and we follow Lin {\it et al.} (1969), Shu (1970) and
Lynden-Bell and Kalnajs (1972), making use of expressions
for the unperturbed epicyclic trajectories of stars (12)
in the equilibrium central field $\Phi_{\mathrm{e}} (r)$. 
In Eqs. (12), $r_0$ is the radius of the
circular orbit, which is chosen so that the constant of areas for this
circular orbit $r_0^2 (d\varphi_0 /dt)$ is equal to the angular
momentum imtegral $M_z= r^2(d\varphi /dt)$, and $v_\perp^2 = v_r^2 +
(2\Omega / \kappa)^2 v_\varphi^2$.  Also, $\varphi_0$ is the position
angle on the circular orbit, $(d\varphi_0 /dt)^2 = (1/r_0)(\partial
\Phi_{\mathrm{e}} / \partial r)_0 = \Omega^2$.  The quantities $\Omega$,
$\kappa$ and $c_r$ are evaluated at $r_0$.  In Eq. (11)
the fact is used that as follows from Eqs. (12)
in a rotating frame the
radial velocity dispersion $c_r$ and the azimuthal velocity dispersion
$c_\varphi$ are connected through $c_r \approx (2\Omega / \kappa)
c_\varphi$.  In the Solar vicinity, $2 \Omega / \kappa \approx 1.7$.
The distribution function $f_{\mathrm{e}}$ has been normalized according 
to $\int_{-\infty}^\infty \int_{-\infty}^\infty f_{\mathrm{e}} d v_r d
v_\varphi = 2\pi (\kappa / 2 \Omega) \int_0^\infty v_\perp d v_\perp
f_{\mathrm{e}}= \sigma_{\mathrm{e}}$, where $\sigma_{\mathrm{e}}$ is the
equilibrium surface density.  Such a distribution function for the
unperturbed system is particularly important because it provides a fit
to observations (Lin {\it et al.}, 1969; Shu, 1970). 

We proceed by applying  the standard procedure of the quasi-linear
(weakly nonlinear) approach (Lifshitz and Pitaevskii, 1981;
Alexandrov {\it et al.}, 1984; Krall and Trivelpiece, 1986) 
and decompose the time
dependent distribution function $f=f_0 ({\bf v},t) + f_1 ({\bf v},t)$
and the gravitational potential $\Phi =\Phi_0 (r,t)+\Phi_1({\bf
r},t)$ with $|f_1/f_0|\ll 1$ and $|\Phi_1 /\Phi_0| \ll 1$ for all
${\bf r}$ and $t$.  The functions $f_1$ and $\Phi_1$ are
oscillating rapidly in space and time, while the functions $f_0$ and
$\Phi_0$ describe the slowly developing ``background" against which
small perturbations develop; $f_0(t=0) \equiv f_{\mathrm{e}}$ and
$\Phi_0(t=0) \equiv \Phi_{\mathrm{e}}$.  The
distribution $f_0$ continues to distort as long as the distribution
is unstable.  Linearizing Eq. (9) and separating
fast and slow varying variables one obtains:
  \begin{equation}
\frac{df_1}{dt} = \frac{\partial \Phi_1}{\partial r}
\frac{\partial f_0}{\partial v_r}
+ \frac{1}{r} \frac{\partial \Phi_1}{\partial \varphi}
\frac{\partial f_0}{\partial v_\varphi} , \label{eq:fast}
  \end{equation}
  \begin{equation}
\frac{\partial f_0}{\partial t} =
{\big\langle} \frac{\partial \Phi_1}{\partial r} \frac{\partial
f_1}{\partial v_r} + \frac{1}{r} \frac{\partial \Phi_1}
{\partial \varphi} \frac{\partial f_1}
{\partial v_\varphi} {\big\rangle} , \label{eq:slow}
  \end{equation}
where $d/dt$ means the total derivative along the star orbit
(12) and $\langle \cdots \rangle$ denotes the time
average over the fast oscillations.  To emphasize it again, we
are concerned with the growth or decay of small perturbations
from an equilibrium state.

In the epicyclic approximation, the partial derivatives in Eqs.
(13) and (14) transform as follows (Lin {\it et al.}, 1969;
Shu, 1970; Morozov, 1980; Griv and Peter, 1996):
   \begin{equation}
\frac{\partial}{\partial v_r} = v_r \frac{\partial}{\partial
{\mathcal{E}}} - \frac{2\Omega}{\kappa} \frac{v_\varphi}{v_\perp^2}
\frac{\partial}{\partial \phi_0}; \quad \frac{\partial}{\partial
v_\varphi} = \left( \frac{2\Omega} {\kappa} \right)^2 v_\varphi
\frac{\partial}{\partial {\mathcal{E}}} + \frac{2\Omega}{\kappa}
\frac{v_r}{v_\perp^2} \frac{\partial}{\partial \phi_0} .
\label{eq:partial}
   \end{equation}

To determine oscillation spectra, let us consider the stability
problem in the lowest WKB approximation; this is accurate
for short wave perturbations only, but qualitatively correctly
even for perturbations with a longer wavelength, of the order of
the disk radius $R$.  In this local WKB approximation in
Eqs. (13) and (14), assuming the weakly
inhomogeneous disk, the perturbation is selected in the form of
a plane wave (in the rotating frame):
   \begin{equation}
f_1, \Phi_1 = \frac{1}{2} \delta f, \delta \Phi \left(
e^{i k_r r + im \varphi - i \omega_* t} + \mathrm{c. c.}
\right) , \label{eq:wkb}
   \end{equation}
where $\delta f$, $\delta \Phi$ are amplitudes that are constant
in space and time, $m$ is the nonnegative azimuthal mode number
(= number of spiral arms),
$\omega_* =\omega - m\Omega$ is the Doppler-shifted
wavefrequency and $|k_r| R \gg 1$ (Lin {\it et al.}, 1969; Shu,
1970; Griv and Peter, 1996).  The solution in such a form
represents a spiral wave with $m$ arms whose shape in the plane
is determined by the relation 
   \begin{equation}
k_r (r - r_0) = - m (\varphi - \varphi_0) .
   \end{equation}
With $\varphi$ increasing in the rotation direction,
we have $k_r > 0$ for trailing spiral patters, which are the most
frequently observed among spiral galaxies.  A change of the sign of
$k_r$ corresponds to changing the sense of winding of the spirals,
i.e. leading ones.  With $m=0$, we have the density waves in the
form of concentric rings that propagate away from the center
when $k_r >0$ or toward the center when $k_r <0$.

In Eq. (13) using the transformation of the derivatives
$\partial /\partial v_r$ and $\partial /\partial v_\varphi$ given
by Eqs. (15), one obtains the solution
   \begin{equation}
f_1= \int_{-\infty}^t d t^\prime {\bf v}_\perp
\frac{\partial \Phi_1}{\partial {\bf r}}
\frac{\partial f_0}{\partial {\mathcal{E}}} , \label{eq:f1}
   \end{equation}
where $f_1 (t^\prime = -\infty) \rightarrow 0$.
In this equation making use of the
time dependence of perturbations in the form of Eq. (16)
and the unperturbed trajectories of stars (12) in
the exponential factor occurring in the formula (18),
it is straightforward to show that
   \begin{equation}
f_1=- \Phi_1 (r_0) \frac{\partial f_0}{\partial
{\mathcal{E}}} \sum_{l=-\infty}^\infty \sum_{n=-\infty}^\infty
l \kappa \frac{e^{i(n-l)(\phi_0 - \zeta)} J_l (\chi) J_n (\chi)}
{\omega_* - l \kappa} , \label{eq:solution}
   \end{equation}
where $J_l (\chi)$ is the Bessel function of the first kind
of order $l$, $\chi =k_*v_\perp /\kappa \sim k_*\rho$,
$k_* =k\{ 1+[(2\Omega /\kappa)^2 - 1] \sin^2 \psi \}^{1/2}$
is the effective wavenumber, $\psi$ is the pitch angle of
perturbations, $\tan \psi = k_\varphi / k_r = m / r k_r$,
and we used the expansion
   \begin{displaymath}
\exp (\pm i b \sin \phi_0) = \sum_{n=-\infty}^\infty
J_n (b) \exp (\pm in\phi_0)
   \end{displaymath}
and the usual Bessel function
recursion relation $J_{l+1} (\chi) + J_{l-1} (\chi) =
(2l/\chi) J_l (\chi)$.  In Eq. (19), the denominator
vanishes when $\omega_* - l \kappa \rightarrow 0$.  This occurs
near corotation ($l=0$) and other resonances ($l
= \pm 1, \pm 2, \cdots$).  The Lindblad resonances occur at
radii where the field $(\partial/\partial {\bf r})\Phi_1$
resonates with the harmonics $l=-1$ (inner
resonance) and $l=1$ (outer resonance) of the epicyclic
(radial) frequency of equilibrium oscillations of stars
$\kappa$.  Clearly, the location of these resonances
depends on the rotation curve and the spiral pattern speed;
the higher the $m$ value, the closer in radius the resonances
are located (Lin {\it et al.}, 1969; Shu, 1970).  
In this paper, only the main part
of the galactic disk is studied which lies sufficiently far from
the resonances: below in all equations $\omega_* -l\kappa \ne 0$.

In the WKB approximation, the linearized Poisson equation (7) is
readily solved to give the improved Lin--Shu expression for the 
perturbed surface density of the two-dimensional disk (Lau and 
Bertin, 1978; Lin and Lau, 1979): 
   \begin{equation}
\sigma_1 = - |k| \Phi_1 / 2\pi G .
   \end{equation}
On the other hand, integrating Eq. (19) over
velocity space $\int dv_r \int
dv_\varphi f_1 = \pi (\kappa / 2\Omega) \int_0^\infty f_1
dv_\perp^2 \equiv \sigma_1$, and equating the result to
the perturbed density given by the improved Lin--Shu solution
(Eq. [20]), the dispersion relation $\omega_* = \omega_* (k_*)$
is obtained:
   \begin{equation}
\frac{k^2 c_r^2}{2 \pi G \sigma_0 |k|} = - \kappa
\sum_{l=-\infty}^\infty l \frac{e^{-x} I_l (x)}{\omega_* - l
\kappa} , \label{eq:dispersion} 
   \end{equation}
where $|\omega_*| \ne l\kappa$, $I_l (x)$ is a Bessel function of
imaginary argument with its argument $x = k_*^2 c_r^2 / \kappa^2 
\approx k_*^2 \rho^2$,  This dispersion relation generalizes 
the Lin--Shu--Kalnajs one for nonaxisymmetric perturbations ($\psi
\ne 0$) propagating in a homogeneous disk excluding the resonance
zones ($\omega_* - l\kappa \ne 0$).  Analyzing Eq. (21), it is
useful to distinguish between two limiting cases: the cases of 
of epicyclic radius that is small compared with wavelength, $x 
\lesssim 1$, and of epicyclic radius that is large compared with
wavelength, $x \gg 1$.
We study the problem when the frequency of disk
oscillations is smaller than the epicyclic frequency.  In the
opposite case of the high perturbation frequencies, $|\omega_*|
> \kappa$, the effect of the disk rotation is negligible and 
therefore not relevant to us.  This is because in this ``rotationless"
case the star motion is approximately
rectilinear on the time and length scales of interest which are
the wave growth/damping periods and wavelengths, 
respectively (cf. Alexandrov {\it et al.}, 1984, p. 113).
Thus, the terms in series (19) and (21) for
which $|l| \ge 2$ may be neglected, and
consideration will be limited to the transparency
region between the turning points in a disk, i.e. between the inner
and outer Lindblad resonances.  The $l=0$ harmonic that defines the
corotation resonance dominates in series. 

Thus, $|\omega_*|$ less than the epicyclic frequency of any disk 
star.  In this case, in Eq. (21) function $\Lambda (x) = \exp (-x)
I_1 (x)$ starts from $\Lambda (0)=0$, reaches a maximum
$\Lambda_{\mathrm{max}}<1$ at $x \approx 0.5$, and then decreases. 
Hence, the growth rate of the instability has a maximum at
$x<1$ (Griv {\it et al.}, 1999b, Fig. 2 in their paper).
In addition, we consider
only low-$m$ perturbations (which are important only for the
problem of spiral structure):
from now on in all equations $m \sim 1$.  

Using the well-known asymptotic expansions of Bessel functions,
the simplified $(|l| \le 1)$ dispersion relation reads
   \begin{equation}
\omega_*^2 = \omega_{\mathrm{J}}^2 , \label{eq:simple}
   \end{equation}
where $\omega_{\mathrm{J}}^2 \approx \kappa^2 -2\pi G\sigma_0|k|F(x)$
is the squared Jeans frequency and $F \approx 2\kappa^2 e^{-x}I_1(x)
/k^2 c_r^2$ is the so-called reduction factor, which takes into
account the fact that the wave field only weakly affects the stars
with high random velocities.  The latter means that the spiral 
structure is seen more distinctly in objects with the lowest
velocity dispersion.  In the limit $x \lesssim 1$, $F (x)
\approx (k_* / k)^2 [1 - x + (3/4) x^2]$ (but,
of course, in order to be appropriate for a WKB wave we consider
the perturbations with $|k_r| r \gg 1$).  In the opposite
limit $x \gg 1$, $F (x) \approx (1/k\rho)^2
[1 - (1/2\pi x )^{1/2}]$.  Different forms of $F (x)$ are given 
by Athanassoula (1984).  The aperiodic Jeans instability
(gravitational collapse) is possible when 
$\omega_*^2 \equiv \omega_{\mathrm{J}}^2 < 0$.
That is, the real part of the frequency of this unstable motion
vanishes in a rotating frame.  The Jeans instability is driven by 
a strong nonresonant interaction of the gravity fluctuations with
the bulk of the particle population, and the dynamics of Jeans
perturbations can be characterized as a fluid-like interaction.
Therefore, such an 
instability, which implies the displacement of macroscopic
portions of a gravitating system, may be also analyzed through
the use of relatively simple hydrodynamic equations (Lau and
Bertin, 1978; Lin and Lau, 1979; Lin and Bertin, 1984).  
In general, the growth rate of the Jeans instability is 
relatively high $|\Im \omega_{\mathrm{J}}| \sim \Omega$;
perturbations with wavelength $\lambda_{\mathrm{J}} \approx
2 \pi c_r / \kappa$ have the fastest growth rate (Griv {\it 
et al.}, 1999a,b; Griv {\it et al.}, 2000a,b).  In the Solar
vicinity of the Galaxy, $\lambda_{\mathrm{J}} = 2-3$ kpc.

Equation (22) has two roots which describe two
branches of oscillations: long-wavelength, $x \lesssim 1$, and
short-wavelength, $x \gg 1$, Jeans branches.  The dispersion
law of low-$m$ Jeans oscillations is
   \begin{equation}
\omega_{* 1,2} = \pm p |\omega_{\mathrm{J}}| . \label{eq:sol}
   \end{equation}
Here $p=1$ for Jeans-stable 
($\omega_{\mathrm{J}}^2 > 0$) perturbations 
and $p=i$ for Jeans-unstable ($\omega_{\mathrm{J}}^2 < 0$) 
perturbations.  

The Jeans perturbations can be stabilized by the random
velocity spread.  Indeed, if one recalls that such unstable
perturbations are possible only when $\omega_{* 1,2}^2 \equiv
\omega_{\mathrm{J}}^2 < 0$, then by using the condition 
$\omega_{\mathrm{J}}^2 \ge 0$ for all possible $k$, from Eq.
(22) a local stability criterion to suppress the instability
of arbitrary Jeans-type gravity perturbations can be easily
obtained (Griv {\it et al.}, 1999a,b; Griv {\it et al.}, 2000b):
   \begin{equation}
c_r \gtrsim
c_{r,\mathrm{crit}} = c_{\mathrm{T}} \{ 1+(2\Omega /\kappa)^2
- 1 ] \sin^2 \psi \}^{1/2} , \label{eq:criterion}
   \end{equation}
where $c_r$ is the radial-velocity dispersion, $c_{r,\mathrm{
crit}}$ is the critical radial velocity dispersion
and usually in the main part of
a disk-shaped galaxy the parameter $2 \Omega / \kappa = 1.5-1.7$.  It
is clear from this local criterion that stability of nonaxisymmetric
Jeans perturbations $(\sin \psi \ne 0)$ in a differentially rotating
disk $(2 \Omega / \kappa > 1)$ requires a larger velocity dispersion
than Toomre's critical value $c_{\mathrm{T}}$. It is only for rigidly
rotating disks ($2 \Omega / \kappa = 1$) and/or for axisymmetric
perturbations ($\sin \psi = 0$) that the critical value of $c_r$ 
is $c_{\mathrm{T}}$.  Hence, the ordinary 
Toomre's velocity dispersion $c_{\mathrm{T}}$
should stabilize only small-scale axisymmetric (radial) perturbations of
the Jeans type.  The differentially rotating disk however is still unstable
against relatively large-scale nonaxisymmetric (spiral) modes with $\psi \ne
0$, in particular very open modes with $\psi \gtrsim 45^\circ$.
According to Polyachenko (1989) and Polyachenko and Polyachenko
(1997) the marginal stability condition
for Jeans perturbations of an arbitrary degree of axial asymmetry
has been available since 1965 (Goldreich and Lynden-Bell, 1965),
though in a slightly masked form.  A relationship exists between
Eq. (24) and what Toomre (Toomre, 1981; Binney and Tremaine, 1987,
p.~375) called ``swing amplification."

Summarizing, for Jeans-stable differentially rotating stellar disks,
the critical value of the radial velocity dispersion must be
greater than (although of the order of) $c_{\mathrm{T}}$.
In accordance with Eq. (24),
   \begin{equation}
c_{\mathrm{M}} \approx (2\Omega / \kappa) c_{\mathrm{T}} \sim
2 c_{\mathrm{T}}
   \end{equation}
might be the ``new" Toomre's velocity dispersion to suppress
all Jeans perturbations in a stellar disk, including the most
dangerous open ones.  Note that the destabilizing effect of tangential
($\psi \ne 0$) gravitational forces (pitch angle dependent effect) was
first considered by Lau and Bertin (1978) and Lin and Lau (1979)
using a much simpler gaseous approach,
and a stability criterion that is similar to Eq. (25) was also derived.
Clearly, the criterion (25) is only approximate one.  In the present
study, we test the validities of this modified criterion for stability
of Jeans perturbations in a self-gravitating, infinitesimally thin and
almost collisionless disk of stars.

Correspondingly, for Jeans-stable differentially rotating stellar disks,
the critical value of the widely used Toomre's instability parameter
$Q = c_{r,\mathrm{crit}} / c_{\mathrm{T}}$ must be greater than 
$2 \Omega / \kappa 
\sim 2$. (Toomre's $Q$-value is a measure of the ratio of thermal
and rotational stabilization to self-gravitation.)  Observations
already demonstated that stellar (and gaseous) disks of a large number
of galaxies, including Milky Way's disk, are near the boundary of
the gravitational stability with $Q$ between 2 and $2 \frac{1}{2}$ 
over a large range of galactic disks (Fridman {\it et al.}, 1991;
Bottema, 1993).

A general impression of how the spectrum of nonaxisymmetric Jeans
perturbations behaves in a homogeneous nonuniformly rotating disk can be
gained from Figure 11, which shows the dispersion curves in the cases of
Jeans-unstable systems ((a) and (b)), a marginally Jeans-stable system
(c) and a Jeans-stable one (d) for values of $l=0,\pm 1,\pm 2,\pm 3$
(as determined on a computer from Eq. [21]).

\begin{figure}
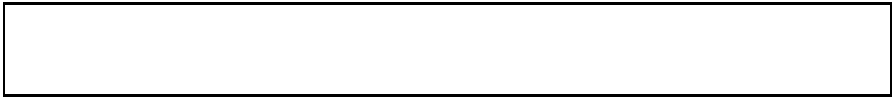

\FigBox{1cm}
\caption{The generalized Lin--Shu--Kalnajs dispersion relation of a
homogeneous stellar disk in the case $2 \Omega / \kappa = \sqrt{2}$
and $|\sin \psi| = 1$ for the different radial velocity dispersion
values: (a) $c_r=0.5 c_{\mathrm{M}}$, (b) $c_r = 0.8 c_{\mathrm{M}}$,
(c) $c_r=c_{\mathrm{M}}$ and (d) $c_r = 1.5 c_{\mathrm{M}}$.
The solid curves represent the real part of the dimensionless
Doppler-shifted wavefrequency $\nu = \omega_*/\kappa$ of
low-frequency ($|\omega_*| < \kappa$)
long-wavelength (1) and short-wavelength (2) Jeans
oscillations we are interested in.  The dashed curves represent the
imaginary part of the dimensionless wavefrequency of low-frequency
vibrations.  The dot-dashed curves represent the wavefrequencies of
additional high-frequency ($|\omega_*| > \kappa$) Jeans modes.}
\label{grivetal11.eps}
\end{figure}

In this figure, the ordinate is the effective wavenumber $k_*$ 
measured in terms of the inverse epicyclic radius $\rho$ and the 
abscissa is $\nu = \omega_*/\kappa$,
i.e. the dimensionless angular frequency at which the stars meet with
the pattern, measured in terms of the epicyclic frequency $\kappa$.
In general, for fixed wavefrequency $\nu$
there are two solutions in $k_* \rho$, comprising a long-wavelength
wave, $k_* \rho \lesssim 1$, and a short-wavelength wave, $k_* \rho
> 1$.  A property of the solution (21) is that in a homogeneous
system the Jeans-stable modes those with $c_r > c_{\mathrm{M}}$ are
separated from each other by frequency intervals where there is no
wave propagation: gaps occur between each harmonic (cf. the Bernstein
modes in a magnetized plasma).

The growth rate of the Jeans instability of a homogeneous stellar
disk $\Im \omega_{\mathrm{J}}$ as determined on a computer from the
expression $\sqrt{2\pi G \sigma_0 |k| F(x)}$ is shown in Figure 12.

\begin{figure}
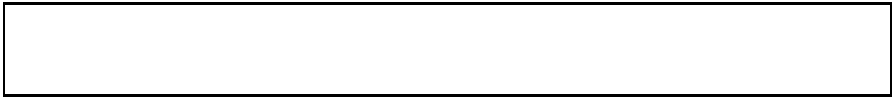

\FigBox{1cm}
\caption{The growth rate of the Jeans instability
of a homogeneous stellar disk (arbitrary units)
in the case $2\Omega /\kappa = \sqrt{2}$, $r_0=8$, $|k_r|=1$, $\rho
=1$ and $|\sin \psi| =1$ for the different values of the azimuthal
mode number $m$.  The growth rates have maxima with respect to the
critical mode number $m_{\mathrm{crit}} \sim 1$. The low-$m$ spiral
modes ($m < 10$ and $m \ne 0$) are more unstable than the radial
ones ($m=0$) and the high-$m$ ones ($m \gtrsim 10$).}
\label{grivetal12.eps}
\end{figure}

As one can see visually in this Figure, the growth rates have maxima
with respect to the critical mode number $m_{\mathrm{crit}} \sim 1$.
It means that of all harmonics of initial perturbation, one
perturbation with the maximum of the growth rate and with
$m=m_{\mathrm{crit}}$ will be formed asymptotically in time.
The low-$m$ spiral modes ($m < 10$ and $m \ne 0$) are more unstable
than the radial ones ($m = 0$) and the high-$m$ ones ($m \gtrsim 10$).
These low-$m$ spiral modes are only important in the problem of
galactic spiral structure because in contrast to the high-$m$ modes,
they do extend essentially over a large range of the galactic disk
(Lin {\it et al.}, 1969; Shu, 1970; Griv and Peter, 1996).

The shape and the number of spiral arms depend on the equilibrium
parameters of a galaxy.  For the Galaxy the most probable pattern
is that of $1-4$ arms, the radial distance between the arms being
$2-4$ kpc.  Such a pattern corresponds to a spiral wave mode with
maximum of growth rate.

Next we substitute the solution (19) into Eq. (14).
Taking into account that the terms $l \ne n$ vanish for axially
symmetric functions $f_0$, after averaging over $\phi_0$ we obtain
the equation for the slow part of the distribution function:
   \begin{equation}
\frac{\partial f_0}{\partial t} = i \pi \sum_{\bf k}
\sum_{l=-\infty}^\infty |\Phi_{1,{\bf k}}|^2 \frac{\partial}
{\partial v_\perp} \frac{k_* \kappa}{v_\perp \chi}
\frac{l^2 J_l^2(\chi)}{\omega_* -l \kappa}
\frac{\partial f_0}{\partial v_\perp} . \label{eq:back}
   \end{equation}

As usual in the weakly nonlinear theory (or weak turbulence theory), 
in order to close the system one must engage an equation for $\Phi_
{1,{\bf k}}$.  Averaging over the fast oscillations, we have
   \begin{equation}
(\partial / \partial t) |\Phi_{1,{\bf k}}|^2
= 2 \Im \omega_* |\Phi_{1,{\bf k}}|^2 , \label{eq:growth}
   \end{equation}
where suffixes $\bf k$ denote the $\bf k$th Fourier component.

Equations (26) and (27) form the closed
system of weakly nonlinear equations for Jeans oscillations of the
rotating homogeneous disk of stars, and describe a diffusion in
velocity space.  The spectrum of oscillations and their growth
rate are given by Eq. (21) and $\approx \sqrt{2\pi G \sigma_0 |k| 
F (x)} \lesssim \Omega$, respectively.  Thus after a time
   \begin{displaymath}
t \sim 1/\Im \omega_* \sim \Omega^{-1}
   \end{displaymath}
the Jeans-unstable disk will assume the form described by Eq.
(17). In the Solar vicinity, $\Omega^{-1} \approx 2 \, 10^8$ yr.
A very important feature of the instability under consideration
is the fact that it is aperiodic (the real part of the
wavefrequency vanishes in a rotating frame we are using).

As an application of the theory we investigate
the relaxation of low frequency and Jeans-unstable,
$|\omega_*| < \kappa$ and $\omega_*^2 < 0$, 
respectively, oscillations in the
homogeneous galactic disk.  Indeed, already in the 1940s it was
observed that in the Solar neighborhood the random velocity
distribution function of stars with an age $t \gtrsim 10^8$ yr
is close to a Schwarzschild distribution --- a set of Gaussian
distributions along each coordinate in velocity space, i.e.
close to equilibrium along each coordinate.  In
addition, older stellar populations have a higher velocity dispersion
than younger ones.
On the other hand, a simple calculation of the relaxation time
of the local disk of the Milky Way due to pairwise star--star
encounters brings the standard value $\sim 10^{14}$ yr (Chandrasekhar,
1960; Binney and Tremaine, 1987, p. 489), which considerably exceeds 
the lifetime of the universe.  According to our approach,
collisionless collective-type relaxation does play a determining 
role in the evolution of stellar populations of the Galactic disk.

Evidently, the unstable fluid-like Jeans oscillations
must influence the distribution function of the main,
nonresonant part of stars in such a way as to hinder the
wave excitation, i.e. to increase the velocity dispersion.
This is because the Jeans instability, being essentially a
gravitational one, tends to be stabilized by random
motions (Griv and Peter, 1996).  Therefore, along with the growth
of the oscillation amplitude, random velocities must increase at the
expense of circular motion, and finally in the disk there can be
established a quasi-stationary distribution so that the Jeans-unstable
perturbations are completely vanishing and only undamped
Jeans-stable waves remain: Jeans instabilities
are thought to heat disks up to velocity dispersion
values of $\gtrsim c_{\mathrm{M}}$.\footnote{In turn,  
the Jeans-stable perturbations are subject to a resonant Landau-type
instability (Griv {\it et al.}, 1999b; Griv {\it et al.}, 2000b).
The resonant interaction of stars with the field of the Jeans-stable
waves will be accompanied by an additional increase in the velocity
dispersion to values $c_r$ greater than it follows from approximate
criterion (25).}

In the following, we restrict ourselves to the most ``dangerous,"
in the sense of the loss of gravitational stability,
long-wavelength perturbations, $\chi^2$ and $x^2 \ll 1$ (see
the explanation after Eq. [21]).  Then in Eqs. (21)
and (26) one can use the expansions $J_1^2 (\chi) 
\approx \chi^2/4$ and $e^{-x} I_1 (x) \approx (1/2)x - (1/2)x^2 +
(5/16)x^3$.  Equation (26) takes the simple form
   \begin{equation}
\frac{\partial f_0}{\partial t} = D \frac{\partial^2 f_0}
{\partial v_\perp^2} , \label{eq:diffusion}
   \end{equation}
where $D=(\pi /16\kappa^2) \sum_{\bf k} k_*^2 \Im \omega_*
|\Phi_{1,{\bf k}}|^2$, and both $\Im \omega_*$ and
$\Phi_{1,{\bf k}}$ are functions of $t$.  As is seen, the
velocity diffusion coefficient for nonresonant stars $D$
is independent of $v_\perp$ (to lowest order).  This is
a qualitative result of the nonresonant character of the
star's interaction with collective aggregates.

By introducing $d\tau/dt = D(t)$ and $d/dt = (d\tau/dt)
(d/d\tau)$ (Alexandrov {\it et al.}, 1984; Krall and Trivelpiece, 
1986), Eqs. (27) and (28) are rewritten as follows:
   \begin{equation}
\frac{\partial f_0}{\partial \tau} - \frac{\partial^2 f_0}
{\partial v_\perp^2} = 0 , \quad \frac{\partial D}{\partial
\tau} = 2 \Im \omega_* , \label{eq:d}
   \end{equation}
which has the solution
   \begin{equation}
f_0 (v_\perp,\tau) = \frac{\mathrm{const}}{\sqrt{\tau}} \exp
\left( - \frac{v_\perp^2}{4 \tau} \right) . \label{eq:relax}
   \end{equation}
(We have taken into account the observations that the
distribution of newly born stars is close to the $\delta$-function
distribution, $f_0 ({\bf v}_\perp,t=0) = \delta ({\bf v}_\perp)$;
Grivnev and Fridman, 1990.)  

Two general physical conclusions can be
deduced from the solution (30).  First, during the
development of the Jeans instability, $\Im \omega_* > 0$, the
Schwarzschild distribution of random velocities (a Gaussian spread
along $v_r,v_\varphi$ coordinates in velocity space) is established.
It sharp contrast to a normal gas with frequent interparticle
collisions, it is the collisionless collective-type interactions
between the stars which transform an arbitrary initial distribution 
into the Maxwellian-like form. 
Secondly, the energy of the oscillation field $\propto \sum_{\bf k}
|\Phi_{1,{\bf k}}|^2$ plays the role of a ``temperature" $T$ in the
nonresonant-particle distribution.  As the perturbation energy
increases, the initially monoenergetic distribution spreads ($f_0$
becomes less peaked), and the effective temperature 
(or $v_\perp^2$) grows with time
(a Gaussian spread increases): $T=2\tau \propto \int D(t)dt\propto
\int \sum_{\bf k}k_*^2 \Im \omega_* |\Phi_{1,{\bf k}}|^2 dt$.
That is, during the development of the Jeans instability the
planar mean-square velocity increases linearly with time.
In Section 4.2 of the present paper, we verify the $v_\perp^2
\propto t$ dependence by $N$-body simulations.

From the above, this nonlinear wave--star interaction increases 
the velocity dispersion of stars in Milky Way's disk after they 
are born (and of particles in Jeans-unstable $N$-body models).
Subsequently, sufficient velocity dispersion prevents Jeans'
instability from occuring.  The ``diffusion" of nonresonant stars
takes place because they gain mechanical (oscillatory) energy as the
instability develops.  The velocity diffusion, however, presumably
tapers off as Jeans stability is approached: the radial velocity
dispersion $c_r$ becomes greater than the critical one $c_{
\mathrm{M}}$.  This is large enough to turn off the Jeans
instability in a stellar disk.  Thus, the true time scale for
relaxation in the Milky Way may be much shorter than its standard
value $\sim 10^{14}$ yr for the classical
Chandrasekhar--Ogorodnikov collisional relaxation; it may
be of the order $(\Im \omega_*)^{-1} \gtrsim \Omega^{-1} \gtrsim 10^9$
yr, i.e. comparable with $10$ periods of the Milky Way rotation in the
Solar vicinity.  The above relaxation time is in fair agreement with
both observations and $N$-body simulations of Milky Way's disk
(Binney and Tremaine, 1987, p. 478; Grivnev and Fridman, 1990; 
Hohl, 1971).

\section{Summary}

The results show that a velocity dispersion given by Toomre's (1964)
criterion will stabilize a disk only against axisymmetric ring-like
gravity perturbations.  However, such disks are unstable against
nonaxisymmetric spiral-like perturbations.  Jeans-unstable spiral
perturbations ``heat" the system; the mean-square random velocity
increases linearly with time.  The results are in agreement with
analytical predictions.

We were able to generate an axisymmetric, gravitationally stable disk.
The initial condition for the axisymmetric stable disk was obtained
analytically and confirmed experimentally: the radial dispersion of
random velocities of stars should be equal (or greater than) to the
modified dispersion at each radii.

\section{Acknowledgements}

This work was performed in part under the auspices of the Israeli
Ministry of Immigrant Absorption, the Israel--U.S. Binational Science
Foundation, the Israel Science Foundation and the Academia Sinica in
Taiwan.  The authors thank Tzi-Hong Chiueh, Arthur D. 
Chernin, Alexei M. Fridman and Shlomi Pistinner for their
interest in the work and valuable suggestions.  In part,  
this work was done while one of the authors (E.~G.) was a Senior
Postdoctoral Fellow at the ASIAA during 1996--1997. E.~G. would like to
thank Yi-Nan Chin, Minho Choi, Yi-Jehng Kuan, Jeremy Lim, Kwok-Yung Lo
and Jun-Hui Zhao for many useful discussions during his two-year stay,
and all the staff of ASIAA for their hospitality.

\end{article}
\end{document}